\documentclass[epj]{webofc}
\usepackage[varg]{txfonts}   % Web of Conferences font
\usepackage{epsf,graphics,amssymb,amsfonts,amsmath,array,feynmf}

\newcommand{\Tr}{{\rm Tr}}

\newcommand{\als}{\alpha_{\rm s}}

\newcommand{\siml}{{\ \lower-1.2pt\vbox{\hbox{\rlap{$<$}\lower6pt\vbox{\hbox{$\sim$}}}}\ }} 
\newcommand{\simg}{{\ \lower-1.2pt\vbox{\hbox{\rlap{$>$}\lower6pt\vbox{\hbox{$\sim$}}}}\ }}

\newcommand{\be}{\begin{equation}} 
\newcommand{\ee}{\end{equation}}
\newcommand{\bea}{\begin{eqnarray}} 
\newcommand{\eea}{\end{eqnarray}}
\newcommand{\beq}{\begin{equation}}
\newcommand{\eeq}{\end{equation}}
\newcommand{\bqa}{\begin{eqnarray}}
\newcommand{\eqa}{\end{eqnarray}}
\def \bx {\mathbf{x}}

\def \bk {\mathbf{k}}

\def \als {\alpha_{\mathrm{s}}}

\def \m2   {\mu^{2 \epsilon}}

\woctitle{QCD@Work 2014}

\begin{document}

\title{Loop functions in thermal QCD}

\author{Antonio Vairo\thanks{\email{antonio.vairo@tum.de}}}
\institute{Physik-Department, Technische Universit\"at M\"unchen,
James-Franck-Str. 1, 85748 Garching, Germany}

\abstract{We discuss divergences of loop functions in thermal QCD and compute perturbatively 
the Polyakov loop, the Polyakov loop correlator and the cyclic Wilson loop. 
We show how these functions get mixed under renormalization.
}

\maketitle

\section{Thermal loop functions}
\label{sec1}
Thermal loop functions are gauge invariant quantities that can be computed by lattice QCD and that are  
relevant for the dynamics of static sources in a thermal bath at a temperature $T$~\cite{McLerran:1981pb}
(for a review, see~\cite{Brambilla:2004wf}). We will focus on three loop functions.

\begin{figure}[ht]
\centering
\includegraphics[width=3cm,clip]{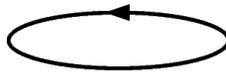}
\caption{Polyakov loop.}
\label{fig1}       
\end{figure}

The {\em Polyakov loop} average in a thermal ensemble at a temperature $T$ is defined as 
\be
\displaystyle P(T)|_R \equiv \frac{1}{d_R}\langle {\Tr} \, L_R \rangle,
\label{PolAv}
\ee
where R is the color representation: $d_A=N^2-1$, $d_F=N$, $N$ is the number of colors, and 
\be
\displaystyle L_R({\bf x}) =  {\rm P} \exp \left( ig \int_0^{1/T}d\tau \, A^0({\bf x},\tau) \right).
\ee
The operator P stands for the path ordering of the color matrices.
A graphical representation is in figure~\ref{fig1}.

\begin{figure}[ht]
\centering
\includegraphics[width=3cm,clip]{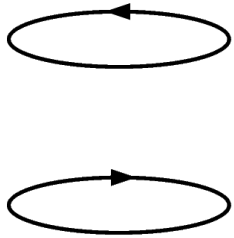}
\caption{Polyakov loop correlator.}
\label{fig2}       
\end{figure}

The {\em Polyakov loop correlator}  is defined as 
\be
P_c(r,T) \equiv \frac{1}{N^2}\langle{\rm Tr} \, L_F^\dagger({\bf 0}) {\rm Tr} \, L_F({\bf r})\rangle,
\label{Polc}
\ee
where ${\bf r}$ is the spatial separation of the two loops.
A graphical representation is in figure~\ref{fig2}. 

\begin{figure}[ht]
\centering
\includegraphics[width=3cm,clip]{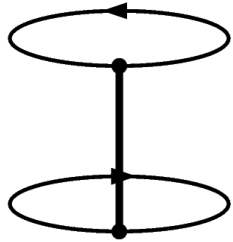}
\caption{Cyclic Wilson loop.}
\label{fig3}       
\end{figure}

The {\em cyclic Wilson loop}  is defined as 
\be
W_c(r,T) \equiv \frac{1}{N}\langle{\rm Tr} \, L_F^\dagger({\bf 0}) U^\dagger(1/T) L_F({\bf r})  U(0)\rangle,
\label{Wcyc}
\ee
where  
\be 
U(1/T) =  {\rm P} \exp \left( ig \int_0^{1}d s \, {\bf r} \cdot {\bf A}(s{\bf r},1/T) \right) = U(0)\,.
\ee
A graphical representation is in figure~\ref{fig3}.

\subsection{Divergences}
\label{sec11}
Loop functions are affected by divergences. These are ultraviolet divergences coming from regions where two or more vertices 
are contracted to one point. In the case of internal vertices, divergences are removed by charge renormalization.
But for loop functions one also gets divergences from the contraction of line vertices 
along the contour. The {\em superficial degree of divergence} is given by $\omega=1-N_{\rm ex}$ at a smooth point 
and $\omega=-N_{\rm ex}$ at a singular point, where $N_{\rm ex}$ is the number of propagators connecting 
the contraction point to uncontracted vertices.

Three type of divergences related to line vertices are possible.
\begin{itemize}
\item[(1)]{~~All vertices are contracted to a smooth point, which leads to a linear divergence.
Linear divergences are proportional to the length of the contour and can be removed by a mass term.} 
\item[(2)]{~~The contraction of vertices to a smooth point leaves an external propagator 
connecting a contracted to an uncontracted vertex: this leads to a logarithmic divergence 
that can be removed by using renormalized fields and couplings~\cite{Dotsenko:1979wb}.}
\item[(3)]{~~All vertices are contracted to a singular point, which gives a logarithmically 
divergent contribution; these are either {\em cusp} or {\em intersection divergences}.}
\end{itemize}

\subsection{Cusps}
\begin{figure}[ht]
 \centering
 \includegraphics[width=0.8\linewidth]{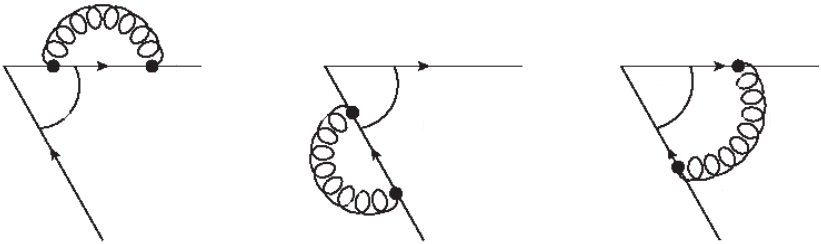}
 \caption{Contributions to a cusp divergence at ${\cal O}(\als)$.}
\label{figcusp}
\end{figure}

One-loop diagrams giving rise to cusp divergences are shown in figure~\ref{figcusp}.
The renormalization constant and the associated \emph{cusp anomalous dimension} at two loops 
can be found in~\cite{Korchemsky:1987wg}. 

A special case is the case of a non-cyclic (time extension smaller than $1/T$) 
rectangular Wilson loop. This has four right-angled cusps. 
The multiplicative renormalization constant in the $\overline{\mathrm{MS}}$-scheme is at one loop 
\be
Z=\exp\left[-{2C_F\als\mu^{-2\varepsilon}}/(\pi\bar{\varepsilon})\right],
\qquad\qquad {1}/\bar{\varepsilon} \equiv {1}/{\varepsilon} -\gamma_E+\ln4\pi\,.
\ee
Cusp divergences are absent in a cyclic Wilson loop.

\subsection{Intersections}
\begin{figure}[ht]
\centering
\includegraphics[width=0.8\linewidth]{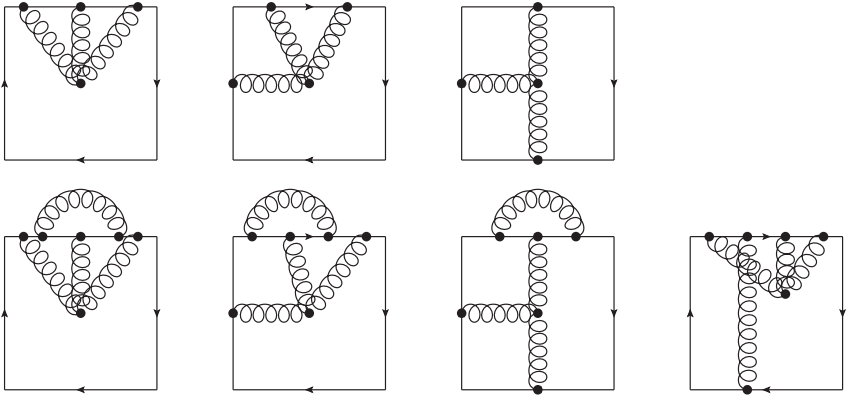}
\caption{Some diagrams contributing to $W_c(r,T)$.
Diagrams in the top row are connected and do not give rise to intersection divergences.
Diagrams in the second row: the first diagram has line vertices only on one quark line, 
so it also contributes to the Polyakov loop, which is finite after charge renormalization; 
the second diagram has vertices on a string and on one quark line and thus cancels through cyclicity; 
the third and fourth diagrams are divergent, because, due to the periodic boundary conditions, 
we can contract the line vertices of the respective one-gluon and three-gluon subdiagrams to
an intersection point.}
\label{figint}
\end{figure}

Divergences appear when all vertices are contracted to an intersection point. 
We restrict here to intersection divergences of a cyclic Wilson loop.
In this case, when one vertex is on the string, if every vertex can be contracted to the intersection,
then the contribution of the diagram cancels because of cyclicity (see~\cite{Berwein:2012mw}).
Moreover, if all vertices are on a quark line, then the diagram contributes 
equally to the Polyakov loop, which is finite after charge renormalization.
Hence a connected diagram cannot give rise to an intersection
divergence, because either we are in one of the situations above,
or it has at least one uncontracted vertex and therefore it is finite. 
Examples of intersection divergent diagrams in a cyclic Wilson loop are the last two diagrams shown in figure~\ref{figint}.

\subsection{Renormalization}
In dimensional regularization, a smooth Wilson loop is finite 
after charge renormalization~\cite{Polyakov:1980ca,Dotsenko:1979wb}. 
If the contour has cusps, additional UV divergences occur. These divergences 
depend only on the angle at the cusp. Such divergences are renormalized through a multiplicative constant. 
We have seen a specific example above.
Intersection divergences arising from an otherwise smooth contour intersecting itself
renormalize instead nontrivially by mixing all possible loops and correlators of loops sharing the same geometry.
In~\cite{Brandt:1981kf}, it was shown that a generic Wilson loop with intersection points connected 
by at most two Wilson lines to other intersection points (angles $\theta_k$) and with cusps (angles $\varphi_l$) 
gets renormalized as
\be
W^{(R)}_{i_1 i_2 \dots i_r}=Z_{i_1j_1}(\theta_1)Z_{i_2j_2}(\theta_2)\cdots Z_{i_rj_r}(\theta_r)
Z(\varphi_1)Z(\varphi_2)\cdots Z(\varphi_s)W_{j_1 j_2 \dots j_r},
\label{generalnormalization}
\ee
where the indices $i_k$ and $j_k$ label the different possible path-ordering prescriptions at the intersection points,
the coupling in $W^{(R)}_{i_1 i_2 \dots i_r}$ is the renormalized coupling and 
the matrices $Z$ are renormalization matrices. 
The loop functions are color-traced and normalized by the number of colours, which 
ensures that all loop functions are gauge invariant.
For some additional remarks on the applicability of the renormalization formula \eqref{generalnormalization} 
we refer to~\cite{Berwein:2013xza}.

\section{Polyakov loop}
\label{sec2}
We consider here the Polyakov loop average defined in \eqref{PolAv}.
It is convenient to compute it in the static gauge, $\partial_0A^0(x)=0$, so that:
\be
L({\bf x}) =\exp\left(\frac{igA^0({\bf x})}{T}\right).
\ee
Propagators may be split into static and non-static components:
\begin{fmffile}{foo}
\begin{eqnarray*}
D_{00}(\omega_n,\bk) \!\!\! &=& \!\!\!
\quad\parbox{30mm}{
\makebox[0truecm]{\phantom b}\put(10,2){
\begin{fmfchar*}(30,0)
\fmfleft{in}
\fmfright{out}
\fmf{dashes}{in,out}
\end{fmfchar*}}}
\;=\; \frac{\delta_{n0}}{\bk^2},
\\
D_{ij}(\omega_n\ne0,\bk) \!\!\! &=& \!\!\!
\quad\parbox{30mm}{
\makebox[0truecm]{\phantom b}\put(10,5){
\begin{fmfchar*}(30,10)
\fmfleft{in}
\fmfright{out}
\fmf{curly}{in,out}
\end{fmfchar*}}}
\;=\; 
\frac{1}{\omega_n^2+\bk^2}\left(\delta_{ij}+\frac{k_ik_j}{\omega_n^2}\right)(1-\delta_{n0}),
\\
D_{ij}(\omega_n=0,\bk) \!\!\! &=& \!\!\!
\quad\parbox{30mm}{
\makebox[0truecm]{\phantom b}\put(10,2){
\begin{fmfchar*}(30,10)
\fmfleft{in}
\fmfright{out}
\fmf{photon}{in,out}
\end{fmfchar*}}}
\;=\;
\frac{1}{\bk^2}\left(\delta_{ij}-(1-\xi)\frac{k_ik_j}{\bk^2}\right)\delta_{n0},
\\
D_{\mathrm{ghost}}(\omega_n,\bk) \!\!\! &=& \!\!\!
\quad\parbox{30mm}{
\makebox[0truecm]{\phantom b}\put(10,2){
\begin{fmfchar*}(30,10)
\fmfleft{in}
\fmfright{out}
\fmf{ghost}{in,out}
\end{fmfchar*}}}
\;=\; \frac{\delta_{n0}}{\bk^2},
\end{eqnarray*}
\end{fmffile}
where  $\omega_n=2\pi n T$, $n \in\mathbb{Z}$, are the bosonic Matsubara frequencies.

Since the temporal component of the gluon self energy is at low momenta: $\Pi_{00}(|\bk|\ll T) = m_D^2 + ...$, 
one has to account also for the {\em Debye mass }, $\displaystyle m_D^2 \equiv \frac{g^2T^2}{3}\left(N+\frac{n_f}{2}\right)$; 
$n_f$ is the number of light quarks.
This constitutes another thermal scale besides the temperature. We will assume the hierarchy
$T \gg m_D$, which is satisfied for weak couplings. 
At the scale $m_D$, the gluon self energies get resummed in the screened temporal-gluon propagator, $1/({\bk^2+m_D^2})$. 
As a consequence, static loops contribute if the flowing momentum is of order $m_D$, 
but vanish (in dimensional regularization) if it is of order $T$.

\begin{figure}[ht]
\makebox[4truecm]{\phantom b}\put(0,0){\includegraphics[width=9cm]{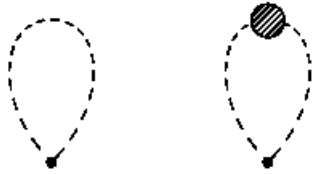}}
\caption{Diagrams contributing to $P(T)|_R$ up to $g^4$ in the static gauge. 
The blob stands for one or more gluon self energy insertions.}
\label{figpolnnlo}
\end{figure}

Up to $g^4$ diagrams contributing to $P(T)|_R$ in the static gauge are shown in figure~\ref{figpolnnlo}.
They give~\cite{Brambilla:2010xn}:
\be
P(T)|_R = 1+\frac{C_R \als }{2}\frac{m_D}{T}+\frac{C_R \als^2}{2}
\left[C_A\left(\ln\frac{m_D^2}{T^2}+\frac{1}{2}\right)-n_f\ln2\right]+\mathcal{O}(g^5),
\label{polnnlo}
\ee
where $C_R$ is the Casimir of the color representation $R$:  $C_F=(N^2-1)/(2N)$, $C_A=N$. 
The logarithm, $\ln{m_D^2}/{T^2}$, signals that an infrared divergence at the scale $T$ has canceled against an ultraviolet 
divergence at the scale $m_D$.

\begin{figure}[ht]
\begin{center}
\includegraphics[width=1.6cm]{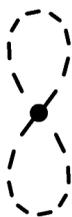}
\end{center}
\caption{Leading diagram contributing to the Casimir scaling violation of the Polyakov loop average in the static gauge.}
\label{figcas}
\end{figure}

In~\cite{Brambilla:2010xn}, also some higher order terms have been calculated.
In particular, non-static modes at the scale $m_D$ contribute with 
\be
\delta P(T)_{\mathrm{NS},\,m_D}=
\frac{3g^4C_R}{4(4\pi)^3}\frac{m_D}{T}\left[\beta_0\ln\left(\frac{\mu }{4\pi T}\right)^2
+2 \beta_0\gamma_E +\frac{11}{3}C_A-\frac{2}{3}n_f\left(4\ln 2-1\right)\right], 
\ee
where $\beta_0$ is the coefficient of the one-loop beta function.
This contribution fixes the renormalization scale of $g^3$ in \eqref{polnnlo} to $\mu \sim 4\pi T$. 
Furthermore, the diagram shown in figure~\ref{figcas} provides the leading contribution 
to the {\em Casimir scaling} violation of the Polyakov loop average 
(i.e. the leading contribution whose color structure is not linear in $C_R$):
\be
\delta P(T) _{\rm Casimir\,viol.}=
\left(3C_R^2-\frac{C_RC_A}{2}\right)\frac{\als^2}{24}\left(\frac{m_D}{T}\right)^2.
\ee
As we remarked above, this contribution, which involves only static loops, 
comes from integrating over momenta scaling like $m_D$.

\subsection{Comparison with the literature}
In 1981, Gava and Jengo obtained in the pure gauge case ($n_f=0$)~\cite{Gava:1981qd}:
\be
P(T)_{\mathrm{GJ}}=
1 +\frac{C_R \als }{2}\frac{m_D}{T}+\frac{C_RC_A\als^2}{2}
\left(\ln \frac{m_D^2}{T^2 }- 2\ln 2+\frac{3}{2}\right)+\mathcal{O}(g^5).
\ee
This result disagrees with \eqref{polnnlo}. The origin of the disagreement can be traced back 
to a missed resummation of the Debye mass in the temporal gluons contributing to the static gluon self energy.

Equation \eqref{polnnlo} agrees instead with the determination of Burnier, Laine and Veps\"al\"ainen, 
who use a dimensionally reduced effective field theory framework in a covariant or Coulomb gauge~\cite{Burnier:2009bk}.

\section{Polyakov loop correlator}
\label{sec3}
\begin{figure}[ht]
\begin{center}
\includegraphics[width=12cm]{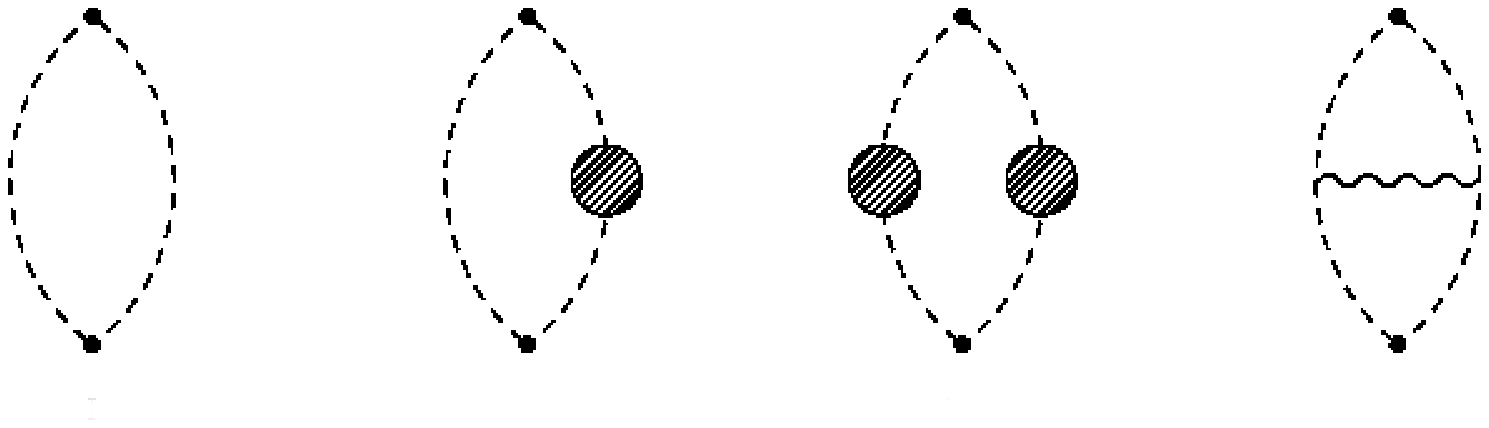}\\
\includegraphics[width=7cm]{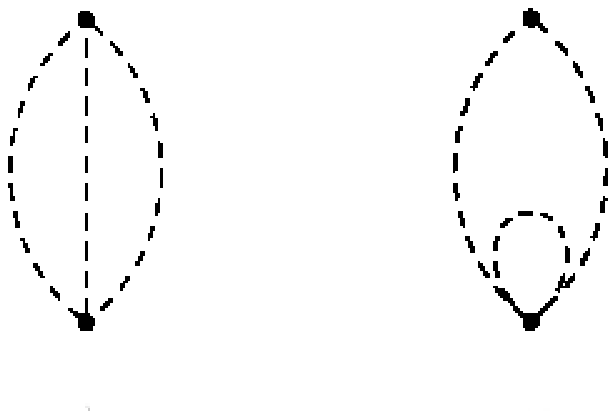}
\end{center}
\caption{Diagrams contributing to $P_c(r,T)$  up to order $g^6(rT)^0$ in the static gauge.}
\label{figPlc}
\end{figure}

In~\cite{Brambilla:2010xn}, the Polyakov loop correlator \eqref{Polc} has been evaluated 
assuming the following hierarchy of scales:
\be
\frac{1}{r}\gg T\gg m_D\gg \frac{g^2}{r}.
\ee
Diagrams contributing to $P_c(r,T)$ up to order $g^6(rT)^0$ are shown in figure~\ref{figPlc}.
They give
\bea
P_c(r,T) &=&  P(T)^2|_F 
+ \frac{N^2-1}{8N^2}\left\{ 
\frac{\als(1/r)^2}{(rT)^2}
-2\frac{\als^2}{rT}\frac{m_D}{T} \right.
+\frac{\als^3}{(rT)^3}\frac{N^2-2}{6N}
\nonumber\\
&&
+ \frac{1}{2\pi }\frac{\als^3}{(rT)^2}\left(\frac{31}{9}C_A-\frac{10}{9}n_f +2\gamma_E\beta_0\right)
+ \frac{\als^3}{rT}\left[
C_A\left(-2 \ln\frac{m_D^2}{T^2} + 2-\frac{\pi^2}{4}\right) + 2n_f\ln 2\right]
\nonumber\\
&&
\left.
+\als^2\frac{m_D^2}{T^2} -\frac{2}{9}\pi \als^3 C_A
\right\}
+\mathcal{O}\left(g^6(rT),\frac{g^7}{(rT)^2}\right).
\label{Plcpert}
\eea

\subsection{Comparison with the literature}
In 1986, Nadkarni calculated the Polyakov loop correlator assuming the hierarchy $T \gg 1/r \sim  m_D$~\cite{Nadkarni:1986cz}.
Whenever the previous results do not involve the hierarchy $rT \ll 1$, 
they agree with Nadkarni's ones expanded for $m_Dr \ll 1$.

Effective field theory approaches for the calculation of the correlator of Polyakov loops 
for the situations $m_D  \simg 1/r$~\cite{Braaten:1994qx} and $T \gg 1/r$~\cite{Nadkarni:1986cz} have been developed since long.
In those situations, the scale $1/r$ was not integrated out, and the Polyakov-loop correlator was
described in terms of dimensionally reduced effective field theories of QCD, while the complexity of the bound-state 
dynamics remained implicit in the correlator. 
Those descriptions are valid for largely separated Polyakov loops when 
the correlator is either screened by the Debye mass, for $m_Dr \sim 1$, or the mass of the
lowest-lying glueball, for $m_Dr \gg 1$.

The Polyakov loop correlator can be put in the form
\be
P_c(r,T) = \frac{1}{N^2}\Bigg[ e^{-{f_{s}(r,T,m_D)}/T} + (N^2-1)e^{-{f_{o}(r,T,m_D)}/T} 
+ {\cal O}\left(\als^3(rT)^4\right)\Bigg],
\ee
where 
$f_{s}$ can be identified with the {\em $Q\bar{Q}$ color-singlet free energy} and 
$f_{o}$ with the  {\em $Q\bar{Q}$ color-octet free energy}.
The color-singlet quark-antiquark potential has been calculated 
in real-time formalism in the same thermodynamical situation considered here in~\cite{Brambilla:2008cx}. 
The comparison of $f_{s}$ with the real-time potential leads to the conclusion that the two cannot be identified 
since the real part of the real-time potential differs from $f_{s}(r,T,m_D)$ by  
$\displaystyle \frac{1}{9} \pi N C_F\als^2rT^2 - \frac{\pi}{36}N^2 C_F\als^3T$.
Moreover, the real-time potential has also an imaginary part that is absent in the free energy.

Jahn and Philipsen have analyzed the gauge structure of the allowed intermediate states
in the correlator of Polyakov loops~\cite{Jahn:2004qr}: the quark-antiquark component, $\varphi$, of an intermediate state  
made of a quark located in $\bx_1$  and an antiquark located in $\bx_2$  should transform as 
$\varphi(\bx_1,\bx_2) \to \Omega(\bx_1) \varphi(\bx_1,\bx_2) \Omega^\dagger(\bx_2)$
under a gauge transformation $\Omega$. 
The decomposition of the Polyakov loop correlator in terms of a color singlet and a 
color octet correlator is in accordance with that result,  
for both a $Q\bar{Q}$ singlet and octet field transform in that way. 
We remark, however, a difference in language:
singlet and octet in $f_s$ and $f_o$ refer to the gauge transformation properties of 
the quark-antiquark fields, while, in~\cite{Jahn:2004qr}, they refer to the gauge 
transformation properties of the physical states. In that last sense   
octet states cannot exist as intermediate states in the correlator of Polyakov loops.

\section{Cyclic Wilson loop}
\label{sec4}
Differently from $P(T)$ and $P_c(r,T)$, the cyclic Wilson loop, $W_c(r,T)$, defined in \eqref{Wcyc}, 
is divergent after charge and field renormalization. This divergence is due to intersection points.
Although it may seem that the cyclic Wilson loop has a continuously infinite number of intersection points
(see figure \ref{fig3}),  one needs to care only about the two endpoints, 
for the Wilson loop contour does not lead to divergences in the other ones.
As a special case of \eqref{generalnormalization}, a cyclic Wilson loop renormalizes as~\cite{Berwein:2012mw}:
\be
 \left(\begin{array}{c} W_c^{(R)} \\ P_c \end{array}\right)=
\left(\begin{array}{cc} Z &  1-Z \\ 0 & 1 \end{array}\right)
\left(\begin{array}{c} W_c \\ P_c \end{array}\right),
\label{reneqWc}
\ee
where 
\be
Z=1+Z_1\als\mu^{-2\varepsilon}+Z_2\left(\als\mu^{-2\varepsilon}\right)^2+ {\cal O}(\als^3).
\ee
The renormalization constant $Z_1$ is given by 
\be
Z_1 =-\frac{C_A}{\pi}\frac{1}{\overline{\varepsilon}}.
\ee
The renormalization constant $Z_2$ reabsorbs all divergences of the type $\als^3/(rT)$ showing 
up in the cyclic Wilson loop, whereas, all other divergences at ${\cal O}(\als^3)$ are reabsorbed 
by $Z_1$ (combined with $P_c(r,T)$ at ${\cal O}(\als^2)$, see \eqref{Plcpert})!

The renormalization group equations read
\be
\left\{
\begin{array}{c}
\displaystyle \mu \frac{\mathrm{d}}{\mathrm{d}\mu} \left(W_c^{(R)}  -P_c\right) = \gamma\, \left(W_c^{(R)}  -P_c\right)
\\
\displaystyle \mu \frac{\mathrm{d}}{\mathrm{d}\mu} \als = - \frac{\als^2}{2\pi}\beta_0 + {\cal O}(\als^3)
\end{array}
\right.,
\ee
where $\gamma$ is the anomalous dimension of $W_c^{(R)} -P_c$:
\be
\gamma \equiv \frac{1}{Z}  \mu \frac{\mathrm{d}}{\mathrm{d}\mu} Z = 2C_A\frac{\als}{\pi} + {\cal O}(\als^2).
\ee
The solution of the renormalization group equations at one-loop is
\be
\left(W_c^{(R)}  -P_c\right)(\mu) = \left(W_c^{(R)}  -P_c\right)(1/r)\;
\left( \frac{\als(\mu)}{\als(1/r)}\right)^{-4C_A/\beta_0}.
\ee

In the $\overline{\mathrm{MS}}$-scheme, up to order $g^4$ and including all terms $\als/(rT) \times (\als \ln\mu r )^n$,
assuming the hierarchy of scales $1/r \gg T \gg m_D \gg g^2/r$, 
the final expression for the cyclic Wilson loop reads~\cite{Berwein:2012mw}
\bea
\ln W_c^{(R)}(r,T;\mu) &=& \frac{C_F\als(1/r)}{rT}
\Biggl\{1+\frac{\als}{4\pi}\left[\left(\frac{31}{9}C_A-\frac{10}{9}n_f\right)
+ 2\, \beta_0 \gamma_E \right]\Biggr.
\notag\\
&&\hspace{1.4cm}\left.+\frac{\als C_A}{\pi}
\left[1+2\gamma_E - 2\ln2 +\sum_{n=1}^\infty\frac{2(-1)^n\zeta(2n)}{n(4n^2-1)}(rT)^{2n}\right]\right\}
\notag\\
&&+\frac{4\pi\als C_F}{T}
\int\frac{\mathrm{d}^3k}{(2\pi)^3}
\left(  e^{i {\bf r}\cdot {\bf k}} - 1 \right)
\left[\frac{1}{\mathbf{k}^2+\Pi_{00}^{(T)}(0,\mathbf{k})}-\frac{1}{\mathbf{k}^2}\right]+C_FC_A\als^2
\notag\\
&&
+ \frac{C_F\als}{rT}\left[\left(\frac{\als(\mu)}{\als(1/r)}\right)^{-4C_A/\beta_0} - 1 \right]
+{\cal O}\left(g^5\right),
\eea
where $\Pi_{00}^{(T)}(0,\mathbf{k})$ is the (known) thermal part of the gluon self energy in Coulomb gauge.

We conclude with some remarks on the renormalization of the cyclic Wilson loop.
First, we notice that, although we have computed $W_c$ for $1/r\gg T\gg m_D \gg g^2/r$, 
the renormalization of $W_c$ reflects its general ultraviolet properties and is not bound to a specific hierarchy. 
In particular, the renormalization equation must hold also at large distances, $r m_D \sim 1$. 
There we have
\be
W_c(r,T) = \,1+ \frac{4\pi C_F\als(\mu)}{T}\frac{e^{-m_Dr}}{4\pi r} 
+ \frac{4 C_FC_A\als^2}{T} \,\frac{e^{-m_Dr}}{4\pi r}\,\frac{1}{\varepsilon} + \dots \,.
\ee
The term $\exp(-m_Dr)/(4\pi r)$ is the Fourier transform of the screened temporal gluon propagator, 
$1/(\mathbf{k}^2+m_D^2)$, and the dots stand for finite or higher-order terms.
Indeed, this expression is renormalized by the same renormalization equation \eqref{reneqWc}
with the same renormalization constant $Z$ computed at short distances.

Finally, we observe that loop functions have, in general, power divergences, which factorize and exponentiate 
to give a factor $\exp\left[\Lambda\, L(C)\right]$, where $L(C)$ is the length of the contour $C$
and $\Lambda$ is some linearly divergent constant. Only in dimensional regularization such linear
divergences are absent, but they would be present in other schemes such as e.g.~lattice regularization~\cite{Polyakov:1980ca}.
An efficient way to calculate the exponent of Wilson loops is the so-called replica trick~\cite{Gardi:2010rn,Gardi:2013ita}.
This consists in calculating the exponent of a Wilson loop $W$, i.e. $\ln\langle W\rangle$, by computing 
the left-hand side of 
\be
\langle W_1\cdot W_2\cdots W_N\rangle = 1+N\ln\langle W\rangle+{\cal O}(N^2),
\ee
where  $W_i$ is the $i$th copy of $W$ in a replicated theory of QCD not interacting with the others.
A~renormalized combination is then~\cite{Berwein:2013xza} 
\be
\exp\bigl[-2\Lambda_F/T-\Lambda_Ar\bigr]\,\times\,Z\,\times\,\bigl(W_c(r,T)-P_c(r,T)\bigr),
\label{renWcPc}
\ee
where $Z$ is now understood in the same renormalization scheme as the linear divergences.

\subsection{Implications for lattice QCD}
\label{sec41}
The renormalization of $W_c$ allows a proper calculation of this quantity on the lattice.
The right quantity to compute is the multiplicatively renormalizable combination $W_c-P_c$
(see \eqref{renWcPc}). A finite quantity is 
\be
\frac{(W_c-P_c)(r,T)}{(W_c-P_c)(r_0,T)} \times \frac{(W_c-P_c)(2r_0-r,T)}{(W_c-P_c)(r_0,T)},
\ee
where $r_0$ is a given distance.

\section*{Acknowledgements} 
I acknowledge financial support from the DFG cluster of excellence 
``Origin and structure of the universe'' (http://www.universe-cluster.de).


\begin{thebibliography}{99}
%\cite{McLerran:1981pb}
\bibitem{McLerran:1981pb} 
  L.~D.~McLerran and B.~Svetitsky,
  %``Quark Liberation at High Temperature: A Monte Carlo Study of SU(2) Gauge Theory,''
  Phys.\ Rev.\ D {\bf 24}, 450 (1981).
  %%CITATION = PHRVA,D24,450;%%

%\cite{Brambilla:2004wf} 
\bibitem{Brambilla:2004wf}
  N.~Brambilla {\it et al.},
  CERN-2005-005, (CERN, Geneva, 2005)
  %``Heavy quarkonium physics,''
  [arXiv:hep-ph/0412158].
  %%CITATION = HEP-PH 0412158;%%:

%\cite{Dotsenko:1979wb}
\bibitem{Dotsenko:1979wb} 
  V.~S.~Dotsenko and S.~N.~Vergeles,
  %``Renormalizability of Phase Factors in the Nonabelian Gauge Theory,''
  Nucl.\ Phys.\ B {\bf 169}, 527 (1980).
  %%CITATION = NUPHA,B169,527;%%

%\cite{Korchemsky:1987wg}
\bibitem{Korchemsky:1987wg} 
  G.~P.~Korchemsky and A.~V.~Radyushkin,
  %``Renormalization of the Wilson Loops Beyond the Leading Order,''
  Nucl.\ Phys.\ B {\bf 283}, 342 (1987).
  %%CITATION = NUPHA,B283,342;%%

%\cite{Berwein:2012mw}
\bibitem{Berwein:2012mw} 
  M.~Berwein, N.~Brambilla, J.~Ghiglieri and A.~Vairo,
  %``Renormalization of the cyclic Wilson loop,''
  JHEP {\bf 1303}, 069 (2013)
  [arXiv:1212.4413 [hep-th]].
  %%CITATION = ARXIV:1212.4413;%%

%\cite{Brandt:1981kf}
\bibitem{Brandt:1981kf} 
  R.~A.~Brandt, F.~Neri and M.~a.~Sato,
  %``Renormalization of Loop Functions for All Loops,''
  Phys.\ Rev.\ D {\bf 24}, 879 (1981).
  %%CITATION = PHRVA,D24,879;%%

%\cite{Berwein:2013xza}
\bibitem{Berwein:2013xza} 
  M.~Berwein, N.~Brambilla and A.~Vairo,
  %``Renormalization of Loop Functions in QCD,''
  Phys.\ Part.\ Nucl.\  {\bf 45}, no. 4, 656 (2014)
  [arXiv:1312.6651 [hep-th]].
  %%CITATION = ARXIV:1312.6651;%%

%\cite{Polyakov:1980ca}
\bibitem{Polyakov:1980ca} 
  A.~M.~Polyakov,
  %``Gauge Fields as Rings of Glue,''
  Nucl.\ Phys.\ B {\bf 164}, 171 (1980).
  %%CITATION = NUPHA,B164,171;%%

%\cite{Brambilla:2010xn}
\bibitem{Brambilla:2010xn} 
  N.~Brambilla, J.~Ghiglieri, P.~Petreczky and A.~Vairo,
  %``The Polyakov loop and correlator of Polyakov loops at next-to-next-to-leading order,''
  Phys.\ Rev.\ D {\bf 82}, 074019 (2010)
  [arXiv:1007.5172 [hep-ph]].
  %%CITATION = ARXIV:1007.5172;%%

%\cite{Gava:1981qd}
\bibitem{Gava:1981qd} 
  E.~Gava and R.~Jengo,
  %``Perturbative Evaluation of the Thermal Wilson Loop,''
  Phys.\ Lett.\ B {\bf 105}, 285 (1981).
  %%CITATION = PHLTA,B105,285;%%

%\cite{Burnier:2009bk}
\bibitem{Burnier:2009bk} 
  Y.~Burnier, M.~Laine and M.~Veps\"al\"ainen,
  %``Dimensionally regularized Polyakov loop correlators in hot QCD,''
  JHEP {\bf 1001}, 054 (2010)
  [Erratum-ibid.\  {\bf 1301}, 180 (2013)]
  [arXiv:0911.3480 [hep-ph]].
  %%CITATION = ARXIV:0911.3480;%%

%\cite{Nadkarni:1986cz}
\bibitem{Nadkarni:1986cz} 
  S.~Nadkarni,
  %``Nonabelian Debye Screening. 1. The Color Averaged Potential,''
  Phys.\ Rev.\ D {\bf 33}, 3738 (1986).
  %%CITATION = PHRVA,D33,3738;%%

%\cite{Braaten:1994qx}
\bibitem{Braaten:1994qx} 
  E.~Braaten and A.~Nieto,
  %``Asymptotic behavior of the correlator for Polyakov loops,''
  Phys.\ Rev.\ Lett.\  {\bf 74}, 3530 (1995)
  [hep-ph/9410218].
  %%CITATION = HEP-PH/9410218;%%

%\cite{Brambilla:2008cx}
\bibitem{Brambilla:2008cx} 
  N.~Brambilla, J.~Ghiglieri, A.~Vairo and P.~Petreczky,
  %``Static quark-antiquark pairs at finite temperature,''
  Phys.\ Rev.\ D {\bf 78}, 014017 (2008)
  [arXiv:0804.0993 [hep-ph]].
  %%CITATION = ARXIV:0804.0993;%%

%\cite{Jahn:2004qr}
\bibitem{Jahn:2004qr} 
  O.~Jahn and O.~Philipsen,
  %``The Polyakov loop and its relation to static quark potentials and free energies,''
  Phys.\ Rev.\ D {\bf 70}, 074504 (2004)
  [hep-lat/0407042].
  %%CITATION = HEP-LAT/0407042;%%

%\cite{Gardi:2010rn}
\bibitem{Gardi:2010rn} 
  E.~Gardi, E.~Laenen, G.~Stavenga and C.~D.~White,
  %``Webs in multiparton scattering using the replica trick,''
  JHEP {\bf 1011}, 155 (2010)
  [arXiv:1008.0098 [hep-ph]].
  %%CITATION = ARXIV:1008.0098;%%

%\cite{Gardi:2013ita}
\bibitem{Gardi:2013ita} 
  E.~Gardi, J.~M.~Smillie and C.~D.~White,
  %``The Non-Abelian Exponentiation theorem for multiple Wilson lines,''
  JHEP {\bf 1306}, 088 (2013)
  [arXiv:1304.7040 [hep-ph]].
  %%CITATION = ARXIV:1304.7040;%%

\end{thebibliography}
\end{document}